\documentclass{ieeetj}
\usepackage{cite}
\usepackage{amsmath,amssymb,amsfonts}
\usepackage{algorithmic}
\usepackage{graphicx,color}
\usepackage{textcomp}
\usepackage{xcolor}
\usepackage{hyperref}
\hypersetup{hidelinks=true}
\usepackage{algorithm,algorithmic}
\usepackage{multirow}
\usepackage[bb=dsserif]{mathalpha}
\def\BibTeX{{\rm B\kern-.05em{\sc i\kern-.025em b}\kern-.08em
    T\kern-.1667em\lower.7ex\hbox{E}\kern-.125emX}}
\AtBeginDocument{\definecolor{tmlcncolor}{cmyk}{0.93,0.59,0.15,0.02}\definecolor{NavyBlue}{RGB}{0,86,125}}

\def\OJlogo{\vspace{-4pt}$<$Society logo(s) and publication title will appear here.$>$}
\def\seclogo{\vspace{10pt}$<$Society logo(s) and publication title will appear here.$>$}

\def\authorrefmark#1{\ensuremath{^{\textbf{#1}}}}

\newcommand\Bbbbone{%
  \ifdefined\mathbbb%
    \mathbbb{1}%
  \else%
    \boldsymbol{\mathbb{1}}%
  \fi}

\begin{document}
\receiveddate{XX Month, XXXX}
\reviseddate{XX Month, XXXX}
\accepteddate{XX Month, XXXX}
\publisheddate{XX Month, XXXX}
\currentdate{XX Month, XXXX}
\doiinfo{XXXX.2022.1234567}

\markboth{LC-PROTONETS: MULTI-LABEL FEW-SHOT LEARNING FOR WORLD MUSIC AUDIO TAGGING}{Papaioannou {et al.}}

\title{LC-Protonets: Multi-Label Few-Shot Learning for World Music Audio Tagging}

\author{Charilaos Papaioannou\authorrefmark{1,2}, Graduate Student Member, IEEE, Emmanouil Benetos\authorrefmark{2}, Senior Member, IEEE and Alexandros Potamianos\authorrefmark{1}, Fellow, IEEE}
\affil{School of ECE, National Technical University of Athens, Greece}
\affil{Centre for Digital Music, Queen Mary University of London, UK}
\corresp{Corresponding author: Charilaos Papaioannou (email: cpapaioan@mail.ntua.gr).}

\begin{abstract}
We introduce Label-Combination Prototypical Networks (LC-Protonets) to address the problem of multi-label few-shot classification, where a model must generalize to new classes based on only a few available examples. Extending Prototypical Networks, LC-Protonets generate one prototype per label combination, derived from the power set of labels present in the limited training items, rather than one prototype per label. Our method is applied to automatic audio tagging across diverse music datasets, covering various cultures and including both modern and traditional music, and is evaluated against existing approaches in the literature. The results demonstrate a significant performance improvement in almost all domains and training setups when using LC-Protonets for multi-label classification. In addition to training a few-shot learning model from scratch, we explore the use of a pre-trained model, obtained via supervised learning, to embed items in the feature space. Fine-tuning improves the generalization ability of all methods, yet LC-Protonets achieve high-level performance even without fine-tuning, in contrast to the comparative approaches. We finally analyze the scalability of the proposed method, providing detailed quantitative metrics from our experiments. The implementation and experimental setup are made publicly available, offering a benchmark for future research.
\end{abstract}


\maketitle

\section{INTRODUCTION}\label{sec:intro}
\IEEEPARstart{H}{umans} are capable of learning from just a few examples \cite{lake_human-level_2015}, highlighting a significant difference between human and typical deep learning at present. Few-shot learning aims to bridge this gap by introducing methods that can learn from limited data \cite{wang_generalizing_2020}. 
This approach is particularly significant for domains and tasks where large-scale data is not readily available, such as in the field of music information retrieval (MIR) when studying under-represented music cultures.

Several systematic approaches have been undertaken in recent years to enrich MIR datasets with music from diverse cultural backgrounds \cite{serra_creating_2014}. 
In most cases, the produced datasets contain tags with various semantic information (e.g., instruments, genre) associated with each music piece. The task of predicting these tags is called automatic music tagging, or auto-tagging \cite{choi_deep_2018, kim2018sample, won_evaluation_2020, lee2017sample}, and constitutes a multi-label classification problem. These tags often exhibit a highly imbalanced distribution, with a small subset occurring frequently while many appear only rarely.

In our previous work \cite{papaioannou_west_2023}, we utilized transfer learning methods to enhance music understanding with a cross-cultural perspective. However, like similar approaches \cite{choi2017transfer, van_den_oord_transfer_2014}, this method is limited to frequent classes. Traditional multi-label classification methods, such as \cite{tsoumakas2011}, which model label correlations by constructing random subsets of labels, are effective but not designed for scenarios where data is minimal and labels are often imbalanced. Few-shot learning \cite{snell_prototypical_2017, vinyals_matching_2017, sung_learning_2018} addresses these limitations by enabling the inclusion of under-represented tags within well-established music domains, where annotated data is often sparse. Furthermore, it supports the exploration of under-represented domains, such as new or emerging music cultures, where annotated datasets are even more limited.

In order to effectively address the automatic audio tagging problem, the adoption of \emph{multi-label} few-shot learning (ML-FSL) methods is essential. While ML-FSL methods have been employed in recent audio classification research \cite{liang_learning_2022}, to the best of our knowledge, they have not yet been applied to MIR datasets. Such a technique can, in turn, be combined with either foundation models \cite{ma2024foundation} or contrastive learning models \cite{spijkervet2021contrastive}, mapping their general-purpose representations to semantic spaces for more comprehensive music analysis.

The goal of this work is to develop an approach to enable multi-label classification for diverse music cultures, in the presence of imbalanced tags and limited data. The main contributions are as follows:
\begin{itemize}
  \item We propose a multi-label few-shot learning method, ``LC-Protonets", that is based on creating multiple prototypes using the union of the power sets of the support samples' labels.
  \item LC-Protonets are applied to the problem of automatic audio tagging for diverse music cultures, and are shown to outperform other multi-label few-shot learning approaches in the literature. Further, the results are consistent across datasets.
  \item We demonstrate the efficacy of a two-step learning methodology aimed at augmenting the inclusion of labels within imbalanced datasets.
  \item The implementation, along with the experimental setup, is provided in a public repository\footnote{\href{https://github.com/pxaris/LC-Protonets}{\texttt{https://github.com/pxaris/LC-Protonets}}} that can serve as a benchmark for similar research works.
\end{itemize}

\section{RELATED WORK}\label{sec:related}

\subsection{AUTOMATIC WORLD MUSIC TAGGING}\label{subsec:autotagging}

In order to describe a piece of music with metadata, various aspects can be considered, including genre, mood, and instrumentation. 
It is a common approach in the MIR field to regard this non-exhaustive list of metadata as tags that characterize a music work \cite{won_music_2021}. The task of predicting the tags of a music piece given its audio signal is referred to as music auto-tagging \cite{choi_deep_2018, kim2018sample, won_evaluation_2020, lee2017sample}, constituting a multi-label classification problem, where an instance can be annotated with a set of multiple labels.

Several architectures have been proposed to address the automatic music tagging problem. Convolutional-based models, such as VGG-ish \cite{hershey_cnn_2017} and Musicnn \cite{pons_musicnn_2019}, are commonly used to extract knowledge from time-frequency input representations, while architectures like SampleCNN \cite{lee_samplecnn_2018} and TCNN \cite{pandey_tcnn_2019} process the raw audio signal. The Audio Spectrogram Transformer (AST) was introduced and used for audio classification in \cite{gong_ast_2021}. In a recent work aimed at building an advanced foundation model capable of performing several tasks, including auto-tagging \cite{won_foundation_2023}, a BERT-style encoder from HuBERT \cite{hsu_hubert_2021} and a Conformer \cite{zhang_pushing_2022} were employed.

Numerous research efforts in the MIR field have incorporated world music datasets \cite{panteli2018computational}. In \cite{demirel_automatic_2018, ganguli_critiquing_2022}, the authors conducted automatic makam recognition, while \cite{sharma_classification_2021} used deep learning models to classify Indian art music. Research in \cite{nikzat_kdc_2022} and \cite{han_finding_2023} used pitch histograms to analyze Iranian dastgāhi and Korean folk music, respectively. A recent study \cite{papaioannou_west_2023} applied auto-tagging to diverse music datasets, but this approach was not designed to handle under-represented tags.

\subsection{FEW-SHOT LEARNING}\label{subsec:few_shot}

Few-shot learning (FSL) methods \cite{snell_prototypical_2017, vinyals_matching_2017, sung_learning_2018} aim to extract knowledge from limited example instances. These methods have been successfully applied to classification tasks across various domains, including computer vision \cite{koch_siamese_nodate, munkhdalai_meta_2017}, natural language processing \cite{joshi_extending_2018, kaiser_learning_2017}, and acoustic signal processing \cite{lake_one-shot_nodate, arik_neural_2018}. In the MIR field, FSL methods have been employed for tasks such as drum transcription \cite{wang_few-shot_2020-1}, source separation \cite{wang_few-shot_2022}, and single instrument recognition \cite{garcia_leveraging_2021}. To the best of our knowledge, this is the first time FSL methods have been applied to multi-label classification within an MIR task, which represents a significant advancement in the field.

The core concept of conventional FSL involves classifying an instance, referred to as the \emph{query item}, based on similarities calculated from class representations derived from a small set of examples called the \emph{support set}. While this framework works well for multi-class tasks, where each instance belongs to a single class, extending it to multi-label classification presents additional challenges, as each instance can be associated with multiple labels.

Several methods have been proposed to address these challenges. In \cite{alfassy_laso_2019}, a technique was introduced to synthesize samples with multiple labels by combining pairs of examples in the feature space. In \cite{simon_meta-learning_2022}, a module was used to predict the number of labels assigned to an item, which allows for multi-label predictions. Attention-based mechanisms were explored in \cite{hu_multi-label_2021}, and a taxonomy hierarchy between tags was utilized in \cite{liang_learning_2022}. Another strategy, ``One-vs.-Rest'' \cite{cheng_multi-label_2019}, reformulates the multi-label problem into multiple binary classification tasks using different classifiers.

While these methods effectively address multi-label classification, they often introduce additional complexity during training, such as the generative module in \cite{alfassy_laso_2019}, the label count module in \cite{simon_meta-learning_2022}, or complex episode formation in \cite{cheng_multi-label_2019}. We argue that such complexities can be simplified or omitted, streamlining the training process and potentially improving performance.

\section{PROPOSED METHOD}\label{sec:method}

In this section, we first describe the Prototypical Networks method and its adaptation to the multi-label setting, followed by the introduction of our proposed method, ``LC-Protonets''.

\subsection{PROTOTYPICAL NETWORKS}\label{subsec:protos}

Prototypical networks \cite{snell_prototypical_2017} are widely used in few-shot learning (FSL) and function by computing a prototype for each class, which represents the average embedding of the support items belonging to that class. Let $S$ denote the support set, consisting of $N \times K$ examples, where $N$ is the number of unique classes (referred to as the N-way) and $K$ is the number of examples per class (referred to as the K-shot). The prototype for a class $c$, denoted as $\mathbf{p}_c$, is computed as the mean of the embedded support examples for that class:
\begin{equation}
\mathbf{p}_c = \frac{1}{K} \sum_{(\mathbf{x}_i, y_i) \in S} f_\theta(\mathbf{x}_i) \cdot \Bbbbone_{y_i = c},
\end{equation}
where $f_\theta(\mathbf{x}_i)$ represents the embedding of input $\mathbf{x}_i$ through a mapping model, and $\Bbbbone_{y_i = c}$ is an indicator function that equals 1 if the label $y_i$ of $\mathbf{x}_i$ belongs to class $c$.

Once the prototypes are computed, a query set $Q$ consisting of unseen examples is used to test the model. Each query item is classified based on the similarity to the prototypes, typically using Euclidean or cosine distance. Specifically, the query sample $\mathbf{x}_q \in Q$ is assigned to the class whose prototype is the closest in the embedding space:
\begin{equation}
\hat{y}_q = \arg\min_{c} d(f_\theta(\mathbf{x}_q), \mathbf{p}_c),
\end{equation}
where $d$ refers to the chosen distance function. During training, cross-entropy loss and a Softmax function over the computed distances are used to learn the embeddings.

\textbf{Episodic learning}: Prototypical networks are trained using episodic learning, progressing through N-way K-shot episodes. In each episode, $N$ classes are randomly sampled, and $K$ support examples are drawn for each class to form the support set $S$. The query set $Q$ consists of additional examples drawn from the same $N$ classes. The model computes the prototypes from the support set, and the loss is calculated based on the classification accuracy of the query set. This episodic approach encourages the model to generalize better in few-shot settings by simulating small training tasks during learning.

\textbf{Extension to multi-label setting}: We adopt the term ``ML-PNs'' (multi-label Prototypical Networks) to refer to the extension of Prototypical Networks for the multi-label setting. This method follows the extension published in \cite{liang_learning_2022}, where it is referred to as ``Baseline''; however, we prefer a more explicit name here to enhance clarity. In this setting, where each sample may belong to multiple classes, each support item contributes to multiple prototypes. Let $\mathbf{y}_i$ be the set of labels for a given sample $\mathbf{x}_i$. For each label $y_{i,j} \in \mathbf{y}_i$, where $j$ ranges from $1$ to the number of classes $N$, the embedding $f_\theta(\mathbf{x}_i)$ is used to update the prototype corresponding to $y_{i,j}$. Consequently, the prototype for each class $c$ is computed by averaging the embeddings of all support examples that belong to class $c$, even if they have additional labels.

In this setting, the Softmax function is replaced by a Sigmoid function, allowing the model to predict multiple labels for each query item. Binary cross-entropy loss is then used to optimize the model:
\begin{equation}
\mathcal{L} = - \sum_{q \in Q} \sum_{c} y_{q,c} \log(\hat{y}_{q,c}) + (1 - y_{q,c}) \log(1 - \hat{y}_{q,c}),
\end{equation}
where $y_{q,c}$ represents the true label for class $c$ for query $q$, and $\hat{y}_{q,c}$ is the predicted probability for that class.

\subsection{LC-PROTONETS}\label{subsec:lc_protonets}

Adapting few-shot learning to the multi-label regime presents a significant challenge, particularly because classes are correlated and each sample may belong to multiple classes. To address this, we propose Label-Combination Prototypical Networks (LC-Protonets), an approach to multi-label classification that extends Prototypical Networks in a simple yet effective way. 

\begin{figure}[t]
    \centering
    \includegraphics[width=0.33\textwidth]{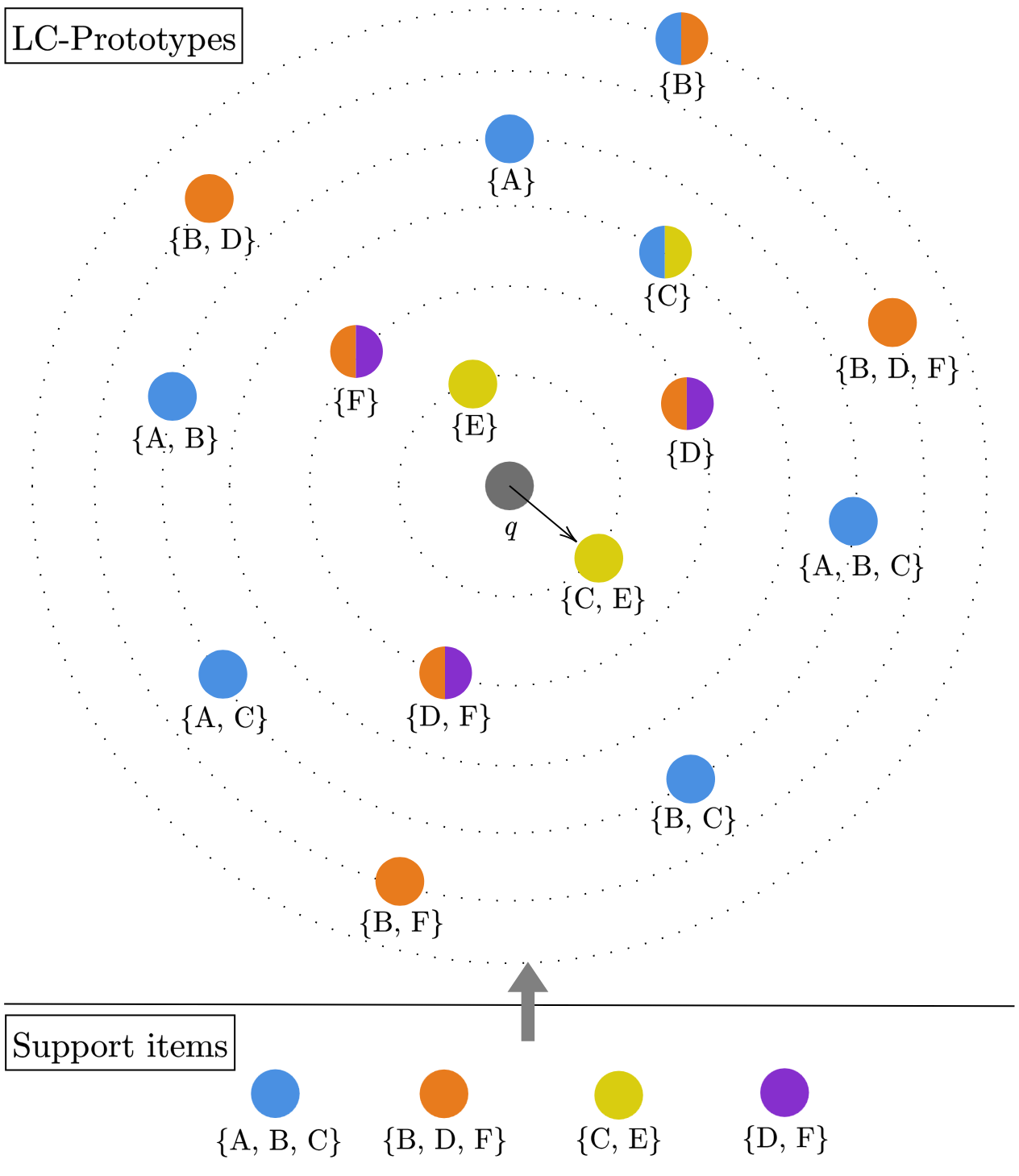}
    \caption{The labels of the support set consisting of four items (bottom) and the LC-classes created from them (top). The concentric circles with the dashed lines pass through LCPs with same representations in the embedding space that are graphed equidistant from the query item $q$.}
    \label{fig:lc_protonets}
\end{figure}

We consider multi-label classification as a problem where every combination of labels is a descriptive label. These combinations are all subsets of the label sets found in the support data, including the full label sets themselves. Hence, a support item $(\mathbf{x}_i, \mathbf{y}_i)$ with $\mathbf{y}_i = \{A, B, C\}$, defines and contributes to all the label combinations derived from the
power set $\mathcal{P}$ of $\mathbf{y}_i$, excluding the empty set:
\begin{equation}
\begin{split}
\mathcal{P}(\{A, B, C\}) = \{\{A\}&, \{B\}, \{C\}, \{A, B\}, \\ & \{A, C\}, \{B, C\}, \{A, B, C\}\}
\end{split}
\end{equation}

Figure \ref{fig:lc_protonets} illustrates an indicative example. The support set $S$ consists of four items, each associated with a distinct set of labels. The set of label combinations, henceforth referred to as \emph{LC-classes}, is defined as the union of the power sets of each $\mathbf{y}_i$ in the support set. Each LC-class is represented by an \emph{LC-Prototype} (LCP), whose representation is computed by averaging the embeddings of the support items that include the corresponding LC-class in their power set, as shown in the color-coded example in the figure.

This approach addresses the multi-label classification problem as a mixture of few-shot and zero-shot learning scenarios. For instance, in Figure \ref{fig:lc_protonets}, the $\{B, D\}$ LC-class does not directly correspond to any support item, but its representation is inferred from items (one in this example) that include it in their label power sets. This introduces a zero-shot learning aspect. Meanwhile, the $\{B, D, F\}$ class has one corresponding item in the support set, enabling a few-shot scenario.

For a query item $q \in Q$, the distances to all LCPs are computed. Figure \ref{fig:lc_protonets} provides a conceptual 2D representation of the embedding space, where concentric circles indicate equal distances from $q$ to different LCPs. In the actual space, the LCPs with identical representations will occupy the same point.

In cases where a query item has equal distances to multiple LCPs, the one representing the largest number of labels is selected. That way, hierarchical relationships and strong correlation between the labels are supported. In the depicted example, both the $\{E\}$ and $\{C, E\}$ LCPs have the minimum distance to $q$, and the query is assigned to the $\{C, E\}$ LC-class. If $E$ is hierarchically subordinate to $C$, the LCPs for $\{E\}$ and $\{C, E\}$ will share the same representation, but the model will consistently select the $\{C, E\}$ class. Even if $C$ and $E$ are not part of a formal hierarchy but they co-occur in the same support items, it is rational to assume strong correlation between them and, again, assign the query to the $\{C, E\}$ class.
 
\textbf{Training phase}: Training is conducted using an episodic ``N-way K-shot'' approach. Here, $N$ refers to the number of active labels (i.e., the number of singleton LC-classes in an episode) and $K$ represents the number of items supporting each singleton LC-class. To prevent oversampling, an item sampled for a singleton class is also counted for any other active classes it belongs to, ensuring that the number of items for each class stays close to $K$.

Given the support items $(\mathbf{x}_i, \mathbf{y}_i) \in S$, the set of all LC-classes $L$ is computed as:
\begin{equation}
L = \bigcup_{(\mathbf{x}_i, \mathbf{y}_i) \in S}^{} \mathcal{P}(\mathbf{y}_i), 
\end{equation}
where $\mathcal{P}(\mathbf{y_i})$ is the power set of the labels of the $i$-th support item, excluding the empty set. The total number of LC-classes is given by the cardinality $|L|$ of the computed set\footnote{The scalability issues of the method in terms of LC-classes are being discussed in detail in Section \ref{sec:results}-\ref{subsec:scalability}.}. 

We denote an LC-class as $L_j$, where $j=1,2,..., |L|$. For each LC-class, one or more support items include it in the power set of their labels, forming the set $S_j \subseteq S$, defined as: $S_j = \{(\mathbf{x}_i, \mathbf{y}_i) \in S \mid L_j \in \mathcal{P}(\mathbf{y}_i)\}.$

The LCP representation $\mathbf{p}_j$ for the corresponding class is computed by averaging the embeddings of the items in $S_j$:
\begin{equation}
\mathbf{p}_j = \frac{1}{|S_j|}\sum_{(\mathbf{x}_i, \mathbf{y}_i) \in S_j}^{}f_\theta(\mathbf{x}_i),
\end{equation}
where $f_\theta$ is the embedding mapping model with $\theta$ trainable parameters.

In each episode, a specified number of query items for each active label is sampled to form the query set $Q$.

Given a query item $\mathbf{x}_i$ with a label set $\mathbf{y}_i$, its initial multi-hot label vector $\mathbf{z}_i \in \{0, 1\}^N$ is constructed such that $\mathbf{z}_i(k) = 1 \quad \text{if} \quad k \in \mathbf{y}_i, \forall k = 1, 2, \ldots, N$. The expanded multi-hot vector $\mathbf{z}_{mH_i} \in \{0, 1\}^{|L|}$ is then constructed by assigning a value of 1 to each of the item's LC-classes: $\mathbf{z}_{mH_i}(j) = 1 \quad \text{if} \quad L_j \in \mathcal{P}(\mathbf{y}_i), \quad \forall j = 1, 2, ..., |L|.$ The loss function is based on the binary cross-entropy:
\begin{equation}
\begin{split}
Loss(\mathbf{x}_i,& \mathbf{z}_{mH_i}, \mathbf{p}) = -\mathbf{z}_{mH_i}\log(\sigma(-d(f_\theta(\mathbf{x}_i), \mathbf{p}))) +\\
&(1-\mathbf{z}_{mH_i})\log(1-\sigma(-d(f_\theta(\mathbf{x}_i), \mathbf{p}))),
\end{split}
\end{equation}
where $d$ is the distance function, $\sigma$ the sigmoid function and $\mathbf{p}$ the LCPs representations. We minimize the loss for all items in the query set $Q$.

\textbf{Inference phase}: At the inference phase, the support set $S$ is created following a similar ``N-way K-shot'' setup used during training. The LC-classes $L$ and their corresponding LCPs representations $\mathbf{p}_j$ (for $j=1,2,...,|L|$) are computed. For a query instance, the distances to all LCPs are calculated, and the instance is assigned to the LC-class represented by the nearest LCP:
\begin{equation}
\hat{\mathbf{y}_i} = \arg\min_{L_j \in L} d(f_\theta(\mathbf{x}_i), \mathbf{p}_j).
\end{equation}

It is important to note that the training phase of LC-Protonets closely follows the extension of Prototypical Networks \cite{snell_prototypical_2017} for the multi-label setting. However, in the latter, only singleton LC-classes are considered in $L$, making it a special case of the LC-Protonets approach. Another difference lies in the inference process, where the probabilities after a Sigmoid layer have to be utilized for classification as opposed to the direct approach adopted by the proposed method.  LC-Protonets transforms the multi-label task to a single-label problem, where every combination of labels $L_j$ is a descriptive label.

% table moved here for formatting reasons
\begin{table}[t]
    \caption{Number of recordings, total tags, and the relative frequency of the $i$\textsuperscript{th} most frequent (and last well-represented) tag for each dataset.}
    \label{table:datasets}
    \begin{center}
        \begin{tabular}{c|c|c|c}
        \hline
        dataset & \# recordings & \# total tags & $i$\textsuperscript{th}: f(\%) \\
        \hline
        MagnaTagATune & 25863 & 188 & 50\textsuperscript{th}: 1.89\% \\
        FMA-medium & 25000 & 151 & 20\textsuperscript{th}: 2.68\% \\
        Lyra & 1570 & 146 & 30\textsuperscript{th}: 4.78\% \\
        Turkish-makam & 5297 & 217 & 30\textsuperscript{th}: 2.83\% \\
        Hindustani & 1204 & 273 & 20\textsuperscript{th}: 2.49\% \\
        Carnatic & 2612 & 283 & 20\textsuperscript{th}: 2.03\% \\
        
        \hline        
        \end{tabular}
    \end{center}
\end{table}

\section{EXPERIMENTS}\label{sec:experiments}

\subsection{DATASETS AND METRICS}\label{subsec:datasets}

We incorporate a range of datasets from both mainstream and world music traditions for our study. For Western music, we use the MagnaTagATune dataset \cite{law_evaluation_2009}, a standard for auto-tagging, and the medium version of FMA \cite{defferrard_fma_2017}. For world music, we utilize the Lyra dataset \cite{papaioannou_dataset_2022}, a collection of Greek folk music, along with three datasets from the CompMusic Corpora \cite{serra_creating_2014}: the Turkish-makam corpus \cite{uyar_corpus_2014, senturk_computational_2016}, which includes recordings from the Turkish tradition, and two datasets of Indian art music, Hindustani and Carnatic \cite{srinivasamurthy_corpora_2014}. 

% results table moved here for formatting reasons
\begin{table*}[ht]
    \caption{Macro-F1 (M-F1) and micro-F1 (m-F1) scores (\%) with 95\% confidence intervals, when aggregating the performance of the three methods over all datasets and training setups, for various ``N-way'' ML-FSL tasks, with a ``3-shot'' approach applied to all of them.}
    \label{table:fsl_aggregated_results}
    \begin{center}
        \begin{tabular}{c|cc|cc|cc|c|c}
        \hline        
        ML-FSL task & \multicolumn{4}{c|}{\textbf{5-way 3-shot}} & \multicolumn{4}{c}{\textbf{15-way 3-shot}} \\
        classes type & \multicolumn{2}{c}{\textit{Base}} & \multicolumn{2}{c|}{\textit{Novel}} & \multicolumn{2}{c}{\textit{Base}} & \multicolumn{2}{c}{\textit{Novel}} \\ \hline
        method / metric & M-F1 & m-F1 & M-F1 & m-F1 & M-F1 & m-F1 & M-F1 & m-F1 \\ \hline
        
        % T1 > upper part
        \textbf{ML-PNs} & \textbf{65.21\textpm4.33} & 66.12\textpm4.41 & 46.32\textpm3.88 & 45.92\textpm3.53 & 39.31\textpm1.66 & 44.23\textpm2.11 & 21.45\textpm1.3 & 21.02\textpm1.22 \\
        \textbf{One-vs.-Rest} & 64.69\textpm4.19 & 65.84\textpm4.16 & 42.69\textpm3.53 & 42.5\textpm3.4 & 35.44\textpm1.64 & 39.4\textpm1.79 & 18.8\textpm1.45 & 18.56\textpm1.41 \\
        \textbf{LC-Protonets (ours)} & 62.6\textpm4.93 & \textbf{66.26\textpm4.87} & \textbf{47.89\textpm6.69} & \textbf{49.34\textpm6.19} & \textbf{42.84\textpm2.71} & \textbf{56.28\textpm2.86} & \textbf{28.5\textpm3.61} & \textbf{31.37\textpm3.74} \\
        
        \hline
        \hline        
        ML-FSL task & \multicolumn{4}{c|}{\textbf{30-way 3-shot}} & \multicolumn{2}{c|}{\textbf{45-way 3-shot}} & \multicolumn{2}{c}{\textbf{60-way 3-shot}} \\
        classes type & \multicolumn{2}{c}{\textit{Base}} & \multicolumn{2}{c|}{\textit{Base \& Novel}} & \multicolumn{2}{c|}{\textit{Base \& Novel}} & \multicolumn{2}{c}{\textit{Base \& Novel}} \\ \hline
        method / metric & M-F1 & m-F1 & M-F1 & m-F1 & M-F1 & m-F1 & M-F1 & m-F1 \\ \hline
        
        % T1 > lower part
        \textbf{ML-PNs} & 29.84\textpm1.14 & 32.57\textpm1.2 & 24.61\textpm0.83 & 29.26\textpm1.27 & 19.74\textpm0.66 & 23.05\textpm1.02 & 17.49\textpm0.62 & 19.45\textpm0.67 \\
        \textbf{One-vs.-Rest} & 25.64\textpm1.21 & 27.81\textpm1.35 & 21.74\textpm0.98 & 25.36\textpm1.29 & 17.06\textpm0.76 & 19.58\textpm1.0 & 14.79\textpm0.68 & 16.17\textpm0.86 \\
        \textbf{LC-Protonets} & \textbf{36.77\textpm2.44} & \textbf{50.77\textpm1.79} & \textbf{31.31\textpm2.14} & \textbf{52.65\textpm1.95} & \textbf{28.09\textpm1.76} & \textbf{50.28\textpm2.06} & \textbf{28.45\textpm1.87} & \textbf{46.48\textpm1.82} \\
        
        \hline        
        \end{tabular}
    \end{center}
\end{table*}

Table \ref{table:datasets} provides details on the number of recordings, total tags, and the relative frequency of the last well-represented label for each dataset. We consider as well-represented the $i$ most frequent tags for each dataset based on their successful inclusion in the supervised learning approach from \cite{papaioannou_west_2023}: $50$ for MagnaTagATune, $30$ for Lyra and Turkish-makam, and $20$ for FMA-medium, Hindustani, and Carnatic. The data preparation for the automatic audio tagging task followed the same process described in \cite{papaioannou_west_2023}. To use these datasets for few-shot learning, we split the labels for training and testing, and we provide those splits to the public repository for reproducibility.

Since our method directly predicts a set of labels without assigning probabilities to individual labels, calculating metrics like Area Under the Curve (AUC) is not straightforward. Therefore, we selected Macro-F1 and Micro-F1 scores for evaluation. F1 score is the harmonic mean of the precision and recall scores and in its Macro- setting, the mean of the individual label scores is calculated. Micro-F1 computes metrics globally by aggregating true positives, false negatives, and false positives across all samples.

\subsection{BACKBONE MODEL}\label{subsec:model}

In few-shot learning, each sample is embedded into a feature space by a backbone model. Given the VGG-ish \cite{simonyan2014very} model's ease of training and proven effectiveness in both supervised learning \cite{papaioannou_west_2023, won_evaluation_2020} and ML-FSL tasks \cite{liang_learning_2022}, we selected it as our backbone model. The architecture consists of a 7-layer Convolutional Neural Network (CNN) with $3\times3$ convolution filters and $2\times2$ max-pooling layers, followed by a couple of fully-connected layers. The model processes log mel-spectrograms as input features.

\subsection{COMPARATIVE APPROACHES}\label{subsec:metrics}

\textbf{ML-PNs}:  Our method is compared to the extension of Prototypical Networks for multi-label classification, referred to as ``ML-PNs''. As described in Section \ref{sec:method}-\ref{subsec:protos}, this approach uses a Sigmoid layer instead of Softmax for classification, and binary cross-entropy loss instead of categorical cross-entropy during training, compared to the standard single-label Prototypical Networks \cite{snell_prototypical_2017}.

\textbf{One-vs.-Rest}: Another comparative approach is the ``One-vs.-Rest'' strategy introduced in \cite{cheng_multi-label_2019}. In this method, the support set is divided into several subsets during training, where each subset focuses on a query’s label along with $N-1$ other classes in an ``N-way K-shot'' format. The goal is to decompose the multi-label problem into multiple binary classification tasks.

\subsection{EXPERIMENTAL SETUP}\label{subsec:experimental_setup}

We split the labels of each dataset into training and testing sets. The training set is used both during the training phase of ML-FSL models and for pre-training the backbone VGG-ish model via supervised learning, while the testing set is used to form the novel classes for evaluation. For the audio recordings, we use the same splits as in previous studies \cite{papaioannou_west_2023, won_evaluation_2020}, with a split ratio of $0.7$, $0.1$, and $0.2$ for the training, validation, and test sets, respectively.

To train the ML-FSL models, we employ three setups: (i) training from scratch with random weight initialization, (ii) full fine-tuning of a pre-trained backbone model, and (iii) fine-tuning only the last layer of the pre-trained backbone model. The model architecture remains the same across all setups, except in the fine-tuning cases where a VGG-ish model pre-trained on the well-represented tags is transferred as the backbone, excluding only the final classification layer.

In few-shot learning, \emph{base} classes are those seen during training, while \emph{novel} classes are unseen. We evaluate ML-FSL models on ``Base'', ``Novel'' and ``Base \& Novel'' classes, with the latter including an equal mix of unseen and seen tags. This allows us to assess how well the model handles both seen and unseen classes during inference. Various values of $N$ (the number of classes) are tested, while $K$ is kept constant (3-shot) across the ``N-way K-shot'' ML-FSL tasks. Importantly, the same \emph{base} classes used for training an ML-FSL model from scratch are also used for pre-training the backbone model via supervised learning, ensuring that the \emph{novel} classes remain unseen for models with a pre-trained backbone.

Cosine distance is used as the distance metric, and all methods are trained using episodic learning with a $10$-way $3$-shot setup. This setup is chosen to accommodate the low-resource nature of music data, as the absolute number of labels differs across domains. For instance, Hindustani, Carnatic, and FMA-medium datasets have only 20 labels in the training split. Additionally, selecting 3 examples per label allows under-represented labels to be included. $50$ episodes are sampled for each epoch and $3$ query items per label are utilized to compute the loss in each one of them.

The validation set is formed by holding out 5 classes from the training set during learning. The Adam optimizer \cite{kingma_adam_2017} is used, and early stopping is applied based on the Macro-F1 score on the validation set. Regarding the input, the audio signal is sampled at 16 kHz, and a 512-point FFT with a 50\% overlap is applied while the Mel bands are set to 128. During training, a random chunk of each audio recording is selected, while during testing, the average embedding of all chunks forms the representation of an instance.

\subsection{TWO-STEP LEARNING METHOD}\label{subsec:two_step}

In imbalanced datasets, it is common to encounter a large number of labels that occur infrequently, leading to a long-tailed label distribution. When training models using supervised learning, a threshold is often set to include only the most frequent categories. To address this limitation, we propose a two-step method that combines supervised and few-shot learning. Unlike ML-FSL setups that fine-tune a pre-trained backbone, this approach requires no fine-tuning.

\begin{figure}[t]
    \centering
    \includegraphics[width=0.48\textwidth]{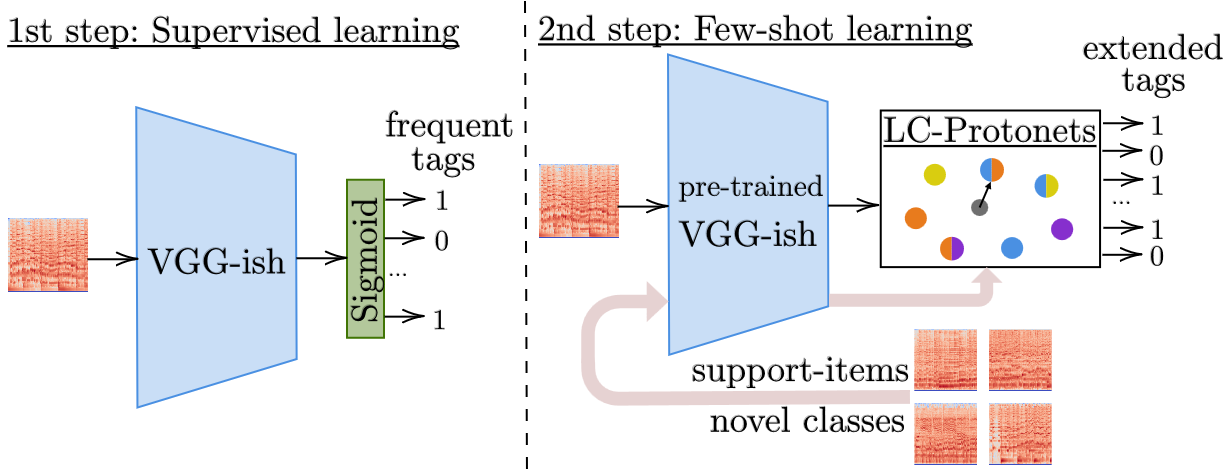}
    \caption{Depiction of the two-step learning method: in the first step, a model is trained on well-represented tags via supervised learning; in the second step, the tag set is extended by applying the LC-Protonets method on the previously trained model, which serves as the backbone.}
    \label{fig:architecture}
\end{figure}

As illustrated in Figure \ref{fig:architecture}, the first step involves training a deep learning model on well-represented tags using supervised learning. The model is optimized with a Sigmoid classification layer and binary cross-entropy loss. In the second step, the pre-trained model is frozen and used as a backbone to map data samples into an embedding space defined by its penultimate layer. We extend the tag set and perform inference on any query item by applying LC-Protonets on top of the pre-trained model. Without additional training, LC-Protonets can classify previously unseen, under-represented labels, including those from the long tail of world music datasets, using just a few examples per label. In our experiments, we use $3$ examples per label.

\begin{table*}[ht]
    \caption{Macro-F1 (M-F1) and micro-F1 (m-F1) scores (\%) with 95\% confidence intervals for a ``30-way 3-shot'' task with ``Base \& Novel'' classes under four conditions: (i) training from scratch, (ii) full fine-tuning of a pre-trained backbone, (iii) fine-tuning only the final layer, and (iv) using a pre-trained backbone without fine-tuning. Rows represent the multi-label few-shot learning methods, and columns correspond to the datasets.}
    \label{table:fsl_results}
    \begin{center}
    \setlength\tabcolsep{2.4pt}
    \resizebox{0.99\textwidth}{!}{%
        \begin{tabular}{ccc|cc|cc|cc|cc|cc}
        \hline        
        dataset & \multicolumn{2}{c|}{\textbf{MagnaTagATune}} & \multicolumn{2}{c|}{\textbf{FMA-medium}} & \multicolumn{2}{c|}{\textbf{Lyra}} & \multicolumn{2}{c|}{\textbf{Turkish-makam}} & \multicolumn{2}{c|}{\textbf{Hindustani}} & \multicolumn{2}{c}{\textbf{Carnatic}} \\ 
        metric & M-F1 & m-F1 & M-F1 & m-F1 & M-F1 & m-F1 & M-F1 & m-F1 & M-F1 & m-F1 & M-F1 & m-F1  \\
        \hline
        method & \multicolumn{12}{c}{\textit{training from scratch}}\\
        
        \hline
        % T2 > scratch
        \textbf{ML-PNs} & 19.89\textpm0.72 & 21.33\textpm1.03 & 18.32\textpm0.54 & 20.04\textpm0.6 & 31.4\textpm1.28 & 37.5\textpm1.71 & 20.29\textpm0.15 & 22.12\textpm0.16 & 18.16\textpm0.68 & 23.89\textpm1.81 & 20.12\textpm0.92 & 30.11\textpm1.39 \\
        \textbf{One-vs.-Rest} & 16.72\textpm0.86 & 17.36\textpm1.13 & 13.73\textpm0.84 & 14.82\textpm0.51 & 29.05\textpm0.41 & 33.9\textpm1.32 & 20.3\textpm0.15 & 22.12\textpm0.14 & 17.86\textpm0.56 & 22.77\textpm1.72 & \textbf{20.23\textpm1.53} & 27.64\textpm1.07 \\
        \textbf{LC-Protonets} & \textbf{21.58\textpm1.56} & \textbf{29.49\textpm2.12} & \textbf{18.91\textpm1.54} & \textbf{34.75\textpm1.73} & \textbf{39.47\textpm4.57} & \textbf{59.82\textpm2.76} & \textbf{21.52\textpm2.95} & \textbf{37.41\textpm2.68} & \textbf{24.71\textpm4.83} & \textbf{50.82\textpm3.49} & 17.96\textpm0.47 & \textbf{54.63\textpm1.4} \\
        
        \hline
        method & \multicolumn{12}{c}{\textit{pre-trained backbone and full fine-tuning}} \\ 
        \hline
        % T2 > full_ft
        \textbf{ML-PNs} & 25.0\textpm0.35 & 27.63\textpm0.57 & 22.08\textpm0.4 & 23.71\textpm0.67 & 35.64\textpm1.31 & 41.72\textpm1.52 & 32.19\textpm1.43 & 32.72\textpm1.46 & 23.1\textpm0.53 & 30.4\textpm0.92 & 22.04\textpm0.79 & 31.41\textpm2.38 \\
        \textbf{One-vs.-Rest} & 19.82\textpm1.09 & 20.96\textpm1.44 & 18.7\textpm0.48 & 19.64\textpm0.88 & 29.02\textpm0.24 & 33.36\textpm0.43 & 26.05\textpm1.42 & 28.08\textpm1.63 & 18.36\textpm0.93 & 23.52\textpm1.66 & 20.72\textpm0.44 & 28.45\textpm0.97 \\
        \textbf{LC-Protonets} & \textbf{33.66\textpm1.41} & \textbf{43.37\textpm2.35} & \textbf{33.37\textpm0.98} & \textbf{48.83\textpm1.36} & \textbf{45.29\textpm2.42} & \textbf{65.99\textpm1.25} & \textbf{38.59\textpm2.45} & \textbf{57.31\textpm2.18} & \textbf{35.07\textpm2.63} & \textbf{59.03\textpm2.33} & \textbf{23.16\textpm1.63} & \textbf{63.69\textpm2.57} \\  
        
        \hline        
        method & \multicolumn{12}{c}{\textit{pre-trained backbone and fine-tuning of the last layer}}\\ 
        \hline
        % T2 > ft_f
        \textbf{ML-PNs} & 24.45\textpm0.59 & 26.94\textpm0.91 & 20.88\textpm0.37 & 22.65\textpm0.46 & 36.72\textpm0.91 & 43.16\textpm0.95 & 28.52\textpm2.66 & 30.73\textpm2.27 & 22.84\textpm0.43 & 31.16\textpm1.43 & 21.38\textpm0.77 & 29.46\textpm2.64 \\
        \textbf{One-vs.-Rest} & 19.86\textpm2.31 & 20.82\textpm2.38 & 19.23\textpm0.44 & 20.52\textpm0.64 & 31.75\textpm1.18 & 37.03\textpm1.65 & 28.95\textpm2.19 & 30.77\textpm1.89 & 20.18\textpm1.78 & 26.29\textpm2.42 & 20.73\textpm0.8 & 28.44\textpm1.36 \\
        \textbf{LC-Protonets} & \textbf{33.5\textpm1.33} & \textbf{43.27\textpm1.98} & \textbf{33.04\textpm1.73} & \textbf{48.68\textpm1.94} & \textbf{47.31\textpm3.16} & \textbf{68.58\textpm1.01} & \textbf{38.52\textpm2.08} & \textbf{57.99\textpm1.48} & \textbf{34.64\textpm2.13} & \textbf{60.04\textpm1.44} & \textbf{23.25\textpm0.69} & \textbf{63.92\textpm1.05} \\
        
        \hline
        method & \multicolumn{12}{c}{\textit{pre-trained backbone without any fine-tuning}}\\
        \hline
        % T2 > pretrained
        \textbf{ML-PNs} & 13.62\textpm0.01 & 14.3\textpm0.01 & 10.58\textpm0.01 & 11.18\textpm0.01 & 28.93\textpm0.09 & 33.25\textpm0.11 & 20.28\textpm0.1 & 22.11\textpm0.12 & 17.49\textpm0.12 & 21.83\textpm0.15 & 20.88\textpm0.07 & 27.76\textpm0.05 \\
        \textbf{One-vs.-Rest} & 13.62\textpm0.01 & 14.31\textpm0.02 & 10.58\textpm0.01 & 11.19\textpm0.01 & 28.9\textpm0.1 & 33.22\textpm0.12 & 20.25\textpm0.06 & 22.07\textpm0.05 & 17.6\textpm0.12 & 21.94\textpm0.18 & 20.91\textpm0.05 & 27.8\textpm0.07 \\
        \textbf{LC-Protonets} & \textbf{33.52\textpm1.19} & \textbf{43.24\textpm2.0} & \textbf{33.73\textpm1.27} & \textbf{49.2\textpm1.87} & \textbf{47.32\textpm3.76} & \textbf{68.95\textpm2.0} & \textbf{37.23\textpm1.71} & \textbf{56.83\textpm0.83} & \textbf{35.09\textpm2.83} & \textbf{59.69\textpm1.81} & \textbf{22.42\textpm0.78} & \textbf{63.66\textpm1.32} \\
        
        \hline
        \end{tabular} 
    }
    \end{center}
\end{table*}

\section{RESULTS}\label{sec:results}

\subsection{ML-FSL TASKS}\label{subsec:ml_fsl}

In Table \ref{table:fsl_aggregated_results}, the aggregated results of the LC-Protonets method and the two comparative approaches are presented. These results were calculated by averaging performance across all datasets and training setups: from scratch, full fine-tuning, and fine-tuning of the last layer. Each experiment was run five times with different random seeds, and the $95\%$ confidence intervals are reported. While label splits remained consistent across runs, different active classes were sampled at each epoch during training, and different support items were selected in each run. We present both macro-F1 and micro-F1 scores for different numbers of labels, ranging from $5$ to $60$. The evaluations were performed on ``Base'' and ``Novel'' classes for smaller numbers of classes, and on ``Base \& Novel'' classes for larger numbers.

\begin{figure}[ht]
    \centering
    \includegraphics[width=0.48\textwidth]{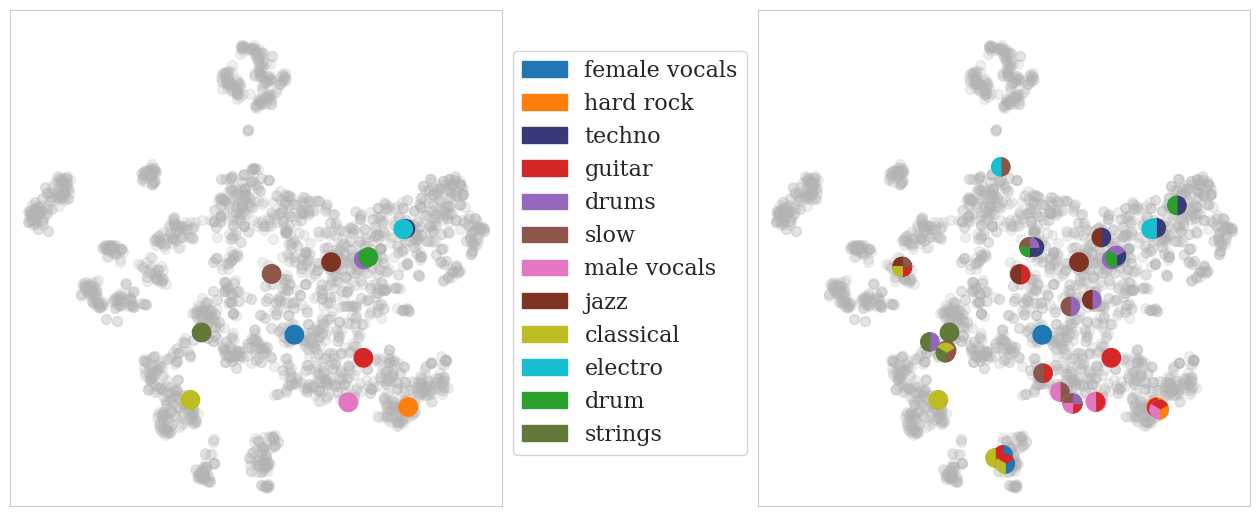}
    \caption{t-SNE visualization of query items (in grey) and prototype embeddings (in distinct colors) for a ``12-way 5-shot'' ML-FSL task on the MagnaTagATune dataset. The left panel shows prototypes generated by the ``ML-PNs'' method (one per class), while the right panel displays those formed using the ``LC-Protonets'' method, where different colors within each prototype indicate the specific label combination it represents.}
    \label{fig:visualilzation}
\end{figure}

LC-Protonets outperformed other methods in nearly all tasks, except for the $5$-way $3$-shot task with base classes, where ML-PNs performed better in terms of macro-F1 score. In the $15$-way $3$-shot task, LC-Protonets showed superior performance  on base classes and widened this gap further when novel classes were used. As the number of classes increased, LC-Protonets demonstrated substantially better performance compared to the other approaches. 

In terms of confidence intervals, we observed wider ranges for few-shot conditions, such as the $5$-way task, due to the random sampling of a small number of active classes in imbalanced datasets. Additionally, LC-Protonets' reliance on support set sampling for deriving LC-classes leads to wider confidence intervals compared to the other methods.

Figure \ref{fig:visualilzation} highlights the differences between the prototypes formed by ML-PNs and LC-Protonets. While ML-PNs create one prototype per class, LC-Protonets populate the embedding space with representations derived from the power sets of the support item labels. We believe this enhanced \emph{positive sampling} of the feature space contributes to the significant performance improvement seen in the results.

Table \ref{table:fsl_results} presents the results of the $30$-way $3$-shot task on ``Base \& Novel'' classes for each dataset and training setup. When training from scratch, the LC-Protonets method showed improvement in all cases except for the macro-F1 evaluation on the Carnatic dataset, where One-vs.-Rest performed better. The difference between LC-Protonets and the comparative approaches was more evident in the micro-F1 scores, as also noted in the aggregated results. 

\begin{figure}[t]
    \centering
    \includegraphics[width=0.42\textwidth]{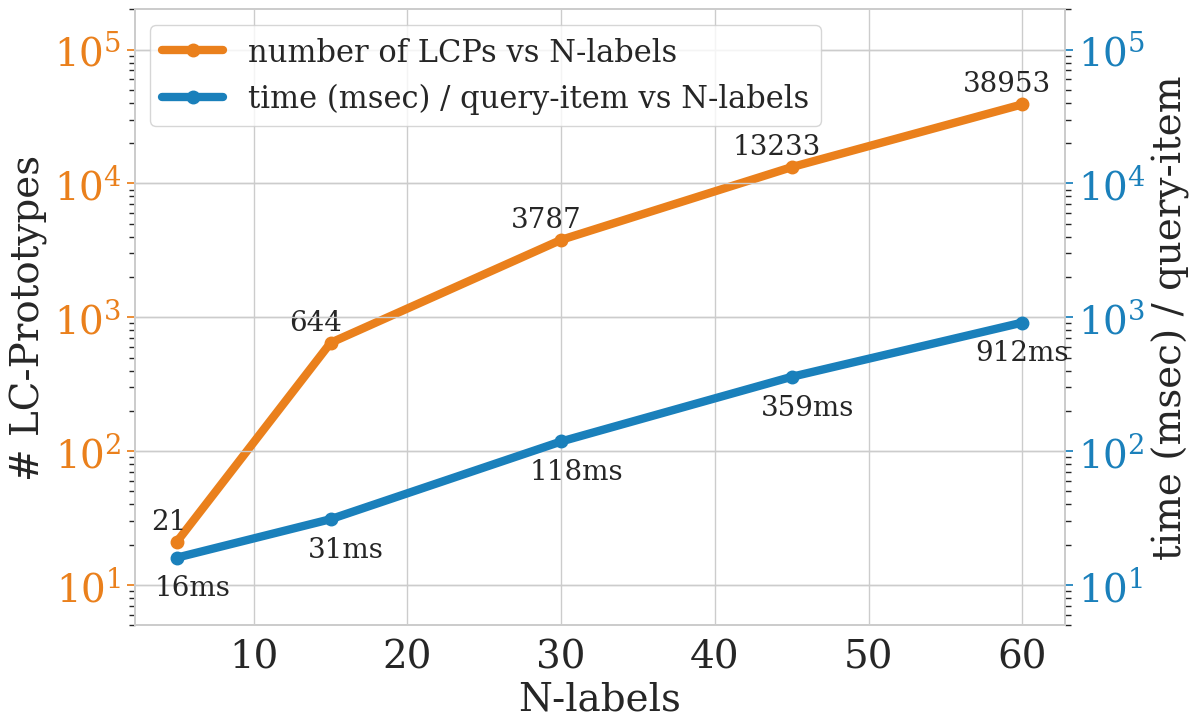}
    \caption{Scalability metrics of the proposed method, averaged across all datasets. The $x$-axis represents the number of labels, the left $y$-axis shows the number of LC-Prototypes, and the right $y$-axis indicates the inference time per item. Both $y$-axes use the same logarithmic scale.}
    \label{fig:scalability}
\end{figure}

When using a pre-trained backbone model followed by full fine-tuning with episodic learning, the performance of all methods significantly improved across all datasets. However, the gap between LC-Protonets and the comparative approaches widened further. For instance, in MagnaTagATune, ML-PNs improved from $19.89\%$ to $25.0\%$, while LC-Protonets increased from $21.58\%$ to $33.66\%$ in macro-F1 score. When only the last layer of the pre-trained model was fine-tuned, performance remained similar across datasets, with the exception of Lyra, where this approach led to slightly better results.

Finally, the last three rows of Table \ref{table:fsl_results} report the performance when a pre-trained backbone model was used without any fine-tuning. Interestingly, LC-Protonets maintained performance levels similar to those seen in the fine-tuning setups. This suggests that the method relies more on the quality of the representations provided by the backbone model than on episodic learning. By contrast, the performance of the comparative approaches dropped significantly, as they encountered challenges with multi-label classification without training, often assigning all labels to all samples. This can be seen in the very narrow confidence intervals of ML-PNs on the FMA-medium dataset, for example.

Comparisons with state-of-the-art methods are not possible due to the lack of ML-FSL results for MIR and these datasets. Moreover, the literature commonly reports ROC-AUC for multi-label tasks instead of Macro-F1. State-of-the-art models also focus on top-N labels, unlike our method which targets under-represented classes.

\begin{table*}[t]
    \caption{Macro-F1 (M-F1) and micro-F1 (m-F1) scores (\%) with 95\% confidence intervals, for the ``VGG-ish'' model on the set of well-represented tags, and for the two-step method ``VGG-ish \& LC-Protonets'' on both the well-represented and extended tag sets for each dataset.}
    \label{table:two_step_learning}
    \begin{center}
    \setlength\tabcolsep{2pt}
    \resizebox{0.99\textwidth}{!}{%
        \begin{tabular}{ccccc|cccc|cccc}
        \hline        
        dataset & \multicolumn{4}{c|}{\textbf{MagnaTagATune}} & \multicolumn{4}{c|}{\textbf{FMA-medium}} & \multicolumn{4}{c}{\textbf{Lyra}} \\ 
        \hline
        \# tags: original / extended & \multicolumn{2}{c}{50} & \multicolumn{2}{c|}{80} & \multicolumn{2}{c}{20} & \multicolumn{2}{c|}{40} & \multicolumn{2}{c}{30} & \multicolumn{2}{c}{60} \\ 
        \hline
        method / metric & M-F1  &  m-F1  & M-F1  &  m-F1  & M-F1  &  m-F1 & M-F1  &  m-F1 & M-F1  &  m-F1 & M-F1  &  m-F1 \\ 
        \hline
        \textbf{VGG-ish} & 26.74\textpm0.63 & 42.29\textpm0.58 & - & - & 36.90\textpm0.76 & 59.61\textpm0.84 & - & - & 30.48\textpm1.23 & 67.14\textpm1.17 & - & - \\
        
        % T3 - upper part
        \textbf{VGG-ish \& LC-Protonets} & 33.09\textpm0.83 & 39.28\textpm1.77 & 26.4\textpm0.26 & 37.31\textpm0.47 & 40.94\textpm2.0 & 53.51\textpm0.73 & 29.12\textpm1.44 & 45.37\textpm1.71 & 47.32\textpm3.76 & 68.95\textpm2.0 & 46.05\textpm2.8 & 69.03\textpm2.21 \\
        
        \hline        
        \hline
        dataset & \multicolumn{4}{c|}{\textbf{Turkish-makam}} & \multicolumn{4}{c|}{\textbf{Hindustani}} & \multicolumn{4}{c}{\textbf{Carnatic}} \\ 
        \hline
        \# tags: original / extended & \multicolumn{2}{c}{30} & \multicolumn{2}{c|}{60} & \multicolumn{2}{c}{20} & \multicolumn{2}{c|}{35} & \multicolumn{2}{c}{20} & \multicolumn{2}{c}{40} \\ 
        \hline
        method / metric & M-F1  &  m-F1  & M-F1  &  m-F1  & M-F1  &  m-F1 & M-F1  &  m-F1 & M-F1  &  m-F1 & M-F1  &  m-F1 \\ 
        \hline
        \textbf{VGG-ish} & 44.95\textpm0.82 & 79.11\textpm0.98 & - & - & 46.07\textpm1.12 & 76.60\textpm1.29 & - & - & 35.49\textpm1.54 & 84.82\textpm1.71 & - & - \\
        
        % T3 - lower part
        \textbf{VGG-ish \& LC-Protonets} & 37.23\textpm1.71 & 56.83\textpm0.83 & 30.07\textpm1.63 & 56.22\textpm1.42 & 40.69\textpm1.83 & 64.38\textpm1.2 & 31.33\textpm2.01 & 58.38\textpm2.41 & 32.1\textpm1.47 & 64.84\textpm1.51 & 18.13\textpm0.6 & 64.25\textpm0.82 \\
        
        \hline        
        \end{tabular}
    }
    \end{center}
\end{table*}

\subsection{TWO-STEP LEARNING METHOD}\label{subsec:two_step_results}

The results of the proposed two-step learning method are shown in Table \ref{table:two_step_learning}. For each dataset, two tag counts are used. The smaller number, such as $20$ for FMA-medium, corresponds to the well-represented tags on which the VGG-ish model was trained, while the larger number, $40$, represents the extended tag set. The performance of both the ``VGG-ish'' model and the ``VGG-ish \& LC-Protonets'' method on the well-represented tags is reported, and for the latter, its performance on the extended tag set is also included.

When examining the macro-F1 scores for both methods on the smaller set of tags, we observe similar performance across most datasets. The key architectural difference between the two approaches is the replacement of the VGG-ish Sigmoid classification layer with the LC-Protonets framework, which classifies an unknown sample to the label combination represented by the nearest LCP. The utilization of the LCPs offers a straightforward way to expand the number of labels. For instance, the tags in MagnaTagATune can be extended from $50$ to $80$, in Hindustani from $20$ to $35$, and doubled for the other datasets. 

There is a relatively small drop in macro-F1 performance as the number of tags increases significantly. For example, in the Turkish-makam dataset, the macro-F1 score drops from $37.23\%$ to $30.07\%$ as the number of tags rises from $30$ to $60$. An exception is the Lyra dataset, where performance on the extended tag set remains nearly identical to the well-represented tags, likely due to stronger correlations between tags in Lyra compared to the other datasets.

\subsection{SCALABILITY}\label{subsec:scalability}

As the LC-classes are derived from the power sets of the sample labels, the number of LC-Prototypes increases significantly as the number of classes $N$ grows. This results in a corresponding increase in inference time, as the distances from all LCPs must be computed for each query item. In Figure \ref{fig:scalability}, the number of classes $N$ is shown on the $x$-axis, while the left $y$-axis represents the number of LC-Prototypes, and the right $y$-axis shows the inference time per query item (in milliseconds), averaged across all datasets. 

When $N$ increases from $5$ to $15$, the number of LCPs rises by a factor of 30, from $21$ to $644$, while inference time increases from $16$ to $31$ milliseconds. As the number of classes continues to grow, the number of LCPs increases substantially, reaching $38,953$ for $60$ tags, and the inference time also rises to $912$ milliseconds. However, the gradient of the increase in inference time is less steep compared to the growth in LCPs. We believe these scalability issues can be mitigated by incorporating optimization techniques, such as skipping the computation of distances from LCPs with identical representations that correspond to different LC-classes (which may be hierarchically related). We plan to explore these methods in future work.

The inference process runs only during testing and not during model training. The average training time across all three methods was under an hour, with no significant differences between them. The trainable parameters amount to $3.66$ million for training from scratch or full fine-tuning, and $262,000$ for fine-tuning only the last layer. The experiments were conducted on an NVIDIA RTX A5000 GPU.

\section{CONCLUSIONS}\label{sec:conclusions}

In this paper, we introduced LC-Protonets, a method for multi-label few-shot classification, and evaluated its performance on automatic audio tagging across diverse music datasets. LC-Protonets consistently outperformed comparative approaches in different domains and ML-FSL tasks, using various training setups. We also explored the impact of fine-tuning a pre-trained backbone model. The results show that utilizing a pre-trained model significantly benefits all ML-FSL methods, and experimenting with both full fine-tuning and fine-tuning of the last layer is recommended.

Additionally, we showed that LC-Protonets can perform well using pre-trained model embeddings without any fine-tuning, unlike the comparative approaches. This led to the development of a two-step learning method, where LC-Protonets can successfully expand the tag set of a dataset by leveraging a model pre-trained on well-represented tags.

However, some limitations remain. The scalability of LC-Protonets, while advantageous in performance, leads to increased inference time as the number of labels grows. This could affect the efficiency of the method when applied to datasets with numerous classes. Additionally, the method's reliance on support set sampling introduces variability, which may impact performance. In future work, we aim on addressing these challenges to further improve robustness and scalability. We will also explore using different architectures and pre-trained models as backbones and attempt experiments in other modalities.

We hope that this work can motivate the development of machine learning methods that can lead to the exploration and understanding of under-represented domains, such as world music cultures, by unlocking the data scarcity problem with a simple yet effective way.

\pagebreak

\bibliographystyle{IEEEtran}
\bibliography{bibtex/bib/oj-sp}

\vfill\pagebreak

\end{document}